\documentclass[onecolumn]{revtex4}
\usepackage{makeidx}
\usepackage{amssymb}
\usepackage{amsmath}
\usepackage{mathrsfs}
\usepackage{graphicx}
\usepackage{dcolumn}
\usepackage{bm}
\usepackage[center]{subfigure}
\usepackage{color}

\begin{document}

\title{Spontaneous symmetry breaking of dual-layer solitons in
spin-orbit-coupled Bose-Einstein condensates}
\author{Zhaopin Chen$^{1}$\thanks{%
Corresponding author.}}
\email{zhaopin.chen@campus.technion.ac.il}
\author{Yongyao Li$^{2}$}
\author{Yan Liu$^{3}$}
\author{Boris A. Malomed$^{4,5}$}
\affiliation{$^1$Physics Department and Solid State Institute, Technion, Haifa 32000,
Israel\\
$^{2}$School of Physics and Optoelectronic Engineering, Foshan University,
Foshan 52800, China\\
$^{3}$Department of Applied Physics, South China Agricultural University,
Guangzhou 510642, China\\
$^{4}$Department of Physical Electronics, School of Electrical Engineering,
Faculty of Engineering, Tel Aviv University, Tel Aviv 69978, Israel\\
$^{5}$Instituto de Alta Investigaci\'{o}n, Universidad de Tarapac\'{a},
Casilla 7D, Arica, Chile}

\begin{abstract}
It is known that stable 2D solitons of the semi-vortex (SV) and mixed-mode
(MM) types are maintained by the interplay of the cubic attractive
nonlinearity and spin-orbit coupling (SOC) in binary Bose-Einstein
condensates. We introduce a double-layer system, in which two binary
condensates, stabilized by the SOC, are linearly coupled by tunneling. By
means of the numerical methods, it is found that symmetric two-layer
solitons undergo the spontaneous-symmetry-breaking (SSB) bifurcation of the
\textit{subcritical} type. The bifurcation produces families of asymmetric
2D solitons, which exist up to the value of the total norm equal to the norm
of the Townes solitons, above which the collapse occurs. This situation
terminates at a critical value of the inter-layer coupling, beyond which the
SSB bifurcation is absent, as the collapse sets in earlier. Symmetric 2D
solitons that are destabilized by the SSB demonstrate dynamical symmetry
breaking, in combination with intrinsic oscillations of the solitons, or
transition to the collapse, if the soliton's norm is sufficiently large.
Asymmetric MMs produced by the SSB instability start spontaneous drift, in
addition to the intrinsic vibrations. Consideration of moving 2D solitons is
a nontrivial problem because SOC breaks the Galilean invariance. It is found
that the system supports moving MMs up to a critical value of the velocity,
beyond which they undergo delocalization.
\end{abstract}

\maketitle

\section{Introduction}

Among many possibilities to realize new physics in Bose-Einstein condensates
(BECs) created in ultracold atomic gases, well-known options are to use them
as testbeds for emulation of various effects in condensed matter physics,
which seem very complex in their original form \cite{Hauke}. In this
context, great interest has been drawn to the experimentally demonstrated
emulation of spin-orbit coupling (SOC) in the binary BEC. In its original
form, SOC originates in physics of semiconductors, as the weakly
relativistic interaction between the electron's magnetic moment and its
motion through the electrostatic field of the ionic lattice \cite%
{Dresselhaus,Bychkov}. Mapping the electron's spinor wave function into the
pseudo-spinor bosonic wave function of the binary condensate under the
action of appropriate laser illumination, SOC is reproduced as the linear
interaction between the pseudospin and momentum of bosonic atoms \cite%
{Lin,Campbell,Ohberg,Galitski,Zhai}. While a majority of experimental works
on the SOC dealt with effectively one-dimensional (1D) SOC settings, the
realization of the SOC in the two-dimensional (2D) BEC was reported too \cite%
{Wu}, which makes it relevant to consider 2D and, eventually, 3D SOC\ states.

The interplay of the linear SOC and usual intrinsic BEC nonlinearity with
the repulsive sign was predicted to produce many nontrivial dynamical
states: vortices \cite{Kawakami,Ramachandhraran,Sakaguchi-Li}, monopoles
\cite{Conduit}, skyrmions \cite{skyrmions,Kawakami-skyrmions},{\ etc.}
Further, the similarity between the Gross-Pitaevskii equations (GPEs) for
the spin-orbit-coupled binary condensate \cite{Achilleos} and the model of
the copropagation of orthogonal polarizations of light in twisted nonlinear
optical fibers~\cite{1991} helps to identify links between the nonlinear
phenomenology in BEC\ and optics. In addition to that, the strong linear
coupling of pseudospin $1/2$ to the atomic momentum connects the physics of
the spinor BEC to graphene physics and its emulation in photonic crystals
\cite{Rechtsman}.

It is natural to expect that solitons in binary condensates with the
attractive sign of the intrinsic nonlinearity may be essentially affected by
SOC. In the general case, a well-known problem is that, in the case of the
cubic self-attraction, 2D fundamental (zero-vorticity, $m=0$) solitons are
unstable in the free space, due to the occurrence of the critical collapse
in this case \cite{Berge,Sulems,Z and K, Fibich}, while solitons carrying
integer vorticity $m$ are subject to a still stronger splitting instability
\cite{PhysD}. In particular, in the 2D case, the GPE with the cubic
self-attraction term gives rise to degenerate families of fundamental Townes
solitons (TSs) \cite{Townes} and their vortical counterparts with $m\geq 1$
\cite{Kruglov2,Kruglov3,Kruglov1}. The degeneracy means that the entire
soliton family with given $m$ shares a single value of the norm, which
separates collapsing and decaying solutions. Therefore, the TSs, that play
the role of separatrices between these two types of the dynamical behavior,
are completely unstable states. In turn, the degeneracy is a consequence of
the specific scale invariance of the cubic GPE in two dimensions.
Peculiarities of the onset of the critical collapse in the 2D SOC system
were studied in Ref. \cite{Mardonov}.

A surprising result, which was first reported in Ref. \cite{Ben-Li}, and
further extended in works \cite{S and M, Salasnich, Evgeny}, is that two
different families of solitons, namely \textit{semi-vortices} (SVs, with
topological charges $m=0$ and $\pm 1$ separately carried by the two
components) and \textit{mixed modes} [MMs, which combine terms with $%
m=\left( 0,-1\right) $ and $m=(0,+1)$ in the components] become \emph{stable}
in the 2D binary system with the self-attraction and linear SOC of the
\textit{Rashba type} \cite{Bychkov,Sherman}. The SV (MM) solitons realize
the ground state of the 2D system when the self-attraction in two components
of the spinor wave function is stronger (weaker) than the cross-attraction
between them. In the special case of the \textit{Manakov's nonlinearity},
with equal strengths of the self- and cross-attraction \cite{Manakov}, the
SOC in the 2D system gives rise to a family of \textit{composite solitons}
(CSs), which continuously connects the SV and MM states with equal values of
the norm and chemical potential, while the ratio of the norms of the two
components, and their angular momenta, vary within the CS family \cite%
{Ben-Li,S and M,we}. The family is dynamically stable against small
perturbations, but is subject to structural instability, as it is destroyed
by any deviation from the Manakov's case.

The explanation of the stabilizing effect of the SOC for the 2D solitons in
free space is provided by the fact that SOC sets up its length scale, which
is inversely proportional to the SOC strength. This fixed length breaks the
above-mentioned scale invariance, thus lifting the norm degeneracy of the
solitons and pushing the norm \emph{below} the threshold necessary for the
onset of the critical collapse. Thus, being protected against the collapse,
the solitons become stable modes that play the role of the system's ground
state, which is missing in the scale-invariant 2D systems with the cubic
self-attraction \cite{Ben-Li} (formally speaking, the collapsing mode is the
ground state in the latter case).

Another topic of general interest in studies of nonlinear-wave systems is
spontaneous symmetry breaking (SSB) between two waveguides (\textit{cores})
with intrinsic nonlinearity (most frequently, it is cubic self-attraction),
coupled by linear terms, which account for tunneling of photons or atoms in
the optical or BEC realization of the system, respectively. For the first
time, the onset of the SSB of 1D solitons in linearly coupled optical
waveguides was predicted in Ref. \cite{Wabnitz}, as the instability of
symmetric solitons (ones with equal components in the coupled cores). The
instability appears at a critical value of the nonlinearity strength, which
is determined by the total norm of the soliton, see below. The analysis of
solutions for the asymmetric 1D solitons, produced by the SSB beyond the
critical point (and, actually, at values of the norm slightly smaller than
the critical one, as the corresponding SSB bifurcation is of the \textit{%
subcritical} type \cite{Iooss}) was developed in detail by means of
numerical simulations and the variational approximation \cite%
{Pare,Maim,Akhmed,Pak}. Experimental observation of the SSB of solitons in
dual-core nonlinear optical fibers was reported only recently \cite{Ignac}.

Similar predictions of the SSB phenomenology were developed for 1D
matter-wave solitons in the double-layer BEC \cite{Arik,Warsaw}. Such a
setting may be implemented in a deep symmetric double-well potential, each
well trapping its layer of the self-attractive condensate; the layers are
coupled by tunneling of atoms across a barrier separating the wells. This
configuration was previously used in theoretical \cite{Milburn, Smerzi} and
experimental \cite{Markus} studies of a related effect, \textit{viz}.,
spontaneous breaking of the \textit{antisymmetry} of spatially odd
continuous-wave states in the \emph{self-repulsive} condensate.

In the case of the 2D double-layer self-attractive BEC, the first problem is
the stabilization of each layer against the above-mentioned critical
collapse. This may be provided by a spatially periodic lattice potential
acting in the layers \cite{Arik}, or by the quadratic (harmonic-oscillator)
trapping potential which is applied in each layer \cite{Luca}. Then, the SSB
gives rise to fundamental and vortical solitons (localized modes), with an
asymmetric distribution of the atomic density between the coupler layers.

As mentioned above, the SOC readily stabilizes 2D solitons of the SV\ and MM
types in the free space, without the use of any external potential. This
property of the spin-orbit-coupled BEC suggests a possibility to predict the
SSB between two components of the 2D solitons in the double-layer setting,
which is the subject of the present work. We find that, similar to the
above-mentioned situation, the solitons of the SV and MM types demonstrate
subcritical bifurcations. However, the existence interval of asymmetric
solitons in terms of the total norm, $N$, above the SSB critical value, $N_{%
\mathrm{cr}}$, is small (somewhat similar to the situation in Ref. \cite%
{Luca}) because the collapse threshold for strongly asymmetric 2D solitons
is close to the TS norm, $N_{\mathrm{TS}}$, while for symmetric solitons it
is $2N_{\mathrm{TS}}$.

The subsequent presentation is structured as follows. The double-layer SOC
system is introduced in Section II, where we also represent its linear
spectrum. This is followed in Section III by a detailed analysis of families
of symmetric and asymmetric 2D solitons in the double-layer system.
Stationary solutions of SV, MM, and composite types are constructed by means
of the accelerated imaginary-time evolution method (AITEM) \cite{AITEM} with
mesh number $N_{x}=N_{y}=256$. Stability of the solutions is identified by
means of systematic simulations of their perturbed evolution, with a total
evolution time $t=400$. The results are reported in a systematic form for
two-layer solitons of the SV and MM types, and briefly for CSs in the
special case of the Manakov's nonlinearity. Section IV addresses moving 2D
solitons. This is a nontrivial problem, as SOC destroys the system's
Galilean invariance, making it impossible to construct moving modes as
boosted copies of quiescent ones.

\section{The double-layer SOC model}

We consider the system modeled by two pairs of spin-orbit-coupled 2D GPEs,
with spatial coordinates $\left( x,y\right) $, for two parallel layers of
the two-component BEC: 

\begin{eqnarray}
i\frac{\partial \phi _{+}}{\partial t} &=&-\frac{1}{2}\nabla ^{2}\phi
_{+}-\left( |\phi _{+}|^{2}+\gamma |\phi _{-}|^{2}\right) \phi _{+}+\lambda
\left( \frac{\partial \phi _{-}}{\partial x}-i\frac{\partial \phi _{-}}{%
\partial y}\right) -\kappa \psi _{+},  \notag \\
i\frac{\partial \phi _{-}}{\partial t} &=&-\frac{1}{2}\nabla ^{2}\phi
_{-}-\left( |\phi _{-}|^{2}+\gamma |\phi _{+}|^{2}\right) \phi _{-}-\lambda
\left( \frac{\partial \phi _{+}}{\partial x}+i\frac{\partial \phi _{+}}{%
\partial y}\right) -\kappa \psi _{-},  \notag \\
&&  \label{phipsi} \\
i\frac{\partial \psi _{+}}{\partial t} &=&-\frac{1}{2}\nabla ^{2}\psi
_{+}-\left( |\psi _{+}|^{2}+\gamma |\psi _{-}|^{2}\right) \psi _{+}+\lambda
\left( \frac{\partial \psi _{-}}{\partial x}-i\frac{\partial \psi _{-}}{%
\partial y}\right) -\kappa \phi _{+},  \notag \\
i\frac{\partial \psi _{-}}{\partial t} &=&-\frac{1}{2}\nabla ^{2}\psi
_{-}-\left( |\psi _{-}|^{2}+\gamma |\psi _{+}|^{2}\right) \psi _{-}-\lambda
\left( \frac{\partial \psi _{+}}{\partial x}+i\frac{\partial \psi _{+}}{%
\partial y}\right) -\kappa \phi _{-},  \notag
\end{eqnarray}%
where $\phi _{\pm }$ and $\psi _{\pm }$ are components of the mean-field
wave functions in the two layers, $\kappa >0$ is the coefficient of the
tunnel coupling between them, and $\gamma \geq 0$ is the relative strength
of the attractive interaction between the components in each layer, while
the self-attraction coefficient is scaled to be $1$. Following Ref. \cite%
{Ben-Li}, SVs and MMs are considered, respectively, for $\gamma =0$ and $%
\gamma =2$. Real coefficient $\lambda $ is the strength of the SOC of the
Rashba type in each layer. The remaining scaling invariance of the system
makes it possible to set $\lambda =1$ (keeping $\kappa $ as a control
parameter), therefore numerical results are presented below for this value.

The total energy of the system includes terms accounting for the kinetic
energy, nonlinearity, SOC, and inter-layer coupling:%
\begin{equation}
E=E_{\mathrm{kin}}+E_{\mathrm{int}}+E_{\mathrm{soc}}+E_{\mathrm{coupl}},
\label{E}
\end{equation}%
where
\begin{equation}
\hspace{-11mm}E_{\mathrm{kin}}=\frac{1}{2}\iint \sum_{+,-}\left( |\nabla
\phi _{\pm }|^{2}+|\nabla \psi _{\pm }|^{2}\right) dxdy,  \label{Ekin}
\end{equation}%
\begin{equation}
\hspace{-11mm}E_{\mathrm{int}}=-\frac{1}{2}\iint \left[ \sum_{+,-}\left(
|\phi _{\pm }|^{4}+|\psi _{\pm }|^{4}\right) +2\gamma \left( |\phi
_{+}|^{2}|\phi _{-}|^{2}+|\psi _{+}|^{2}|\psi _{-}|^{2}\right) \right] dxdy,
\label{Eint}
\end{equation}%
\begin{equation}
\hspace{-11mm}E_{\mathrm{soc}}=\lambda \iint \left[ \phi _{+}^{\ast }\left(
\left( \phi _{-}\right) _{x}-i\left( \phi _{-}\right) _{y}\right) +\psi
_{+}^{\ast }\left( \left( \psi _{-}\right) _{x}-i\left( \psi _{-}\right)
_{y}\right) \right] dxdy+\mathrm{c.c.},  \label{Esoc}
\end{equation}%
\begin{equation}
\hspace{-11mm}E_{\mathrm{coupl}}=-\kappa \sum_{+,-}\iint \phi _{\pm }^{\ast
}\psi _{\pm }dxdy+\mathrm{c.c.},  \label{Ecoup}
\end{equation}%
with $\mathrm{c.c.}$standing for the complex-conjugate expression.

Looking for a solution to the linearized form of Eq. (\ref{phipsi}) in the
form of%
\begin{equation}
\phi _{\pm },\psi _{\pm }\sim \exp \left( -i\mu t+p_{x}x+p_{y}y\right)
\label{linear}
\end{equation}%
leads to the set of four branches of the relation between chemical potential
$\mu $ and wavenumbers $p_{x,y}$ defined in Eq. (\ref{linear}):%
\begin{equation}
\mu =\frac{1}{2}p^{2}\pm \lambda p-\kappa ,\mu =\frac{1}{2}p^{2}\pm \lambda
p+\kappa ,  \label{dispersion}
\end{equation}%
where $p=\sqrt{p_{x}^{2}+p_{y}^{2}}$. This result determines the \textit{%
semi-infinite gap}, in terms of $\mu $, which is not covered by Eq. (\ref%
{dispersion}), hence it may be populated by solitons:%
\begin{equation}
\mu <\mu _{\min }\equiv -\frac{1}{2}\lambda ^{2}-\kappa ,  \label{mu-min}
\end{equation}%
where the minimum value of $\mu (p)$ is attained at $p=\lambda $.

Stationary states with chemical potential $\mu $ are looked for as solutions
to Eq. (\ref{phipsi}) in the form of

\begin{equation}
\phi _{\pm }(x,y,t)=\Phi _{\pm }(x,y)e^{-i\mu t},\psi _{\pm }(x,y,t)=\Psi
_{\pm }(x,y)e^{-i\mu t},  \label{stat_phipsi}
\end{equation}%
where stationary wave functions satisfy equations%
\begin{eqnarray}
\mu \Phi _{+} &=&-\frac{1}{2}\nabla ^{2}\Phi _{+}-\left( |\Phi
_{+}|^{2}+\gamma |\Phi _{-}|^{2}\right) \Phi _{+}+\lambda \left( \frac{%
\partial \Phi _{-}}{\partial x}-i\frac{\partial \Phi _{-}}{\partial y}%
\right) -\kappa \Psi _{+},  \notag \\
\mu\Phi _{-} &=&-\frac{1}{2}\nabla ^{2}\Phi _{-}-\left( |\Phi
_{-}|^{2}+\gamma |\Phi _{+}|^{2}\right) \Phi _{-}-\lambda \left( \frac{%
\partial \Phi _{+}}{\partial x}+i\frac{\partial \Phi _{+}}{\partial y}%
\right) -\kappa \Psi _{-},  \notag \\
&&  \label{PhiPsi} \\
\mu\Psi _{+} &=&-\frac{1}{2}\nabla ^{2}\Psi _{+}-\left( |\Psi
_{+}|^{2}+\gamma |\Psi _{-}|^{2}\right) \Psi _{+}+\lambda \left( \frac{%
\partial \Psi _{-}}{\partial x}-i\frac{\partial \Psi _{-}}{\partial y}%
\right) -\kappa \Phi _{+},  \notag \\
\mu \Psi _{-} &=&-\frac{1}{2}\nabla ^{2}\Psi _{-}-\left( |\Psi
_{-}|^{2}+\gamma |\Psi _{+}|^{2}\right) \Psi _{-}-\lambda \left( \frac{%
\partial \Psi _{+}}{\partial x}+i\frac{\partial \Psi _{+}}{\partial y}%
\right) -\kappa \Phi _{-},  \notag
\end{eqnarray}
The stationary states are characterized by the total norm,

\begin{equation}
N=N_{\phi }+N_{\psi }\equiv N_{\phi _{+}}+N_{\phi _{-}}+N_{\psi
_{+}}+N_{\psi _{-}}=\int \int \left( |\Phi _{+}|^{2}+|\Phi _{-}|^{2}+|\Psi
_{+}|^{2}+|\Psi _{-}|^{2}\right) dxdy.  \label{Norm}
\end{equation}%
The SSB in the present system is quantified by means of the norm difference
between the layers,

\begin{equation}
\theta =\frac{|N_{\phi }-N_{\psi }|}{N_{\phi }+N_{\psi }}.  \label{ratio}
\end{equation}

In the single-layer system ($\kappa =0,~\psi _{\pm }=0$), the family of SV
solitons with $\gamma <1$ is completely stable. It exists in the interval of
\begin{equation}
0<N<N_{\mathrm{TS}}\approx 5.85,  \label{NTS}
\end{equation}%
which corresponds to $\mu <-\lambda ^{2}/2$ [see Eq. (\ref{mu-min})], where $%
N_{\mathrm{TS}}$ is the norm of the single-component TS. In the limit of $%
\mu \rightarrow -\infty $, the two-component SV degenerates into a
single-component TS \cite{Ben-Li,Evgeny}. The MM solitons are unstable at $%
\gamma <1$.

In the case of $\gamma >1$, the SVs are unstable, while the MM family is
completely stable in the single layer$.$ This family exists in the interval
of norms
\begin{equation}
0<N<2N_{\mathrm{TS}}/(\gamma +1).  \label{MM-Townes}
\end{equation}%
In the limit of $\mu \rightarrow -\infty $, the MM state turns into a
symmetric two-component TS, which corresponds to the largest norm in Eq. (%
\ref{MM-Townes}).

There exist two isomers of the SV-type solitons, with right- and left-handed
chiralities, which are defined, respectively, by vorticity sets $\left(
S_{+},S_{-}\right) =\left( 0,+1\right) $ and $\left( \bar{S}_{+},\bar{S}%
_{-}\right) =\left( -1,0\right) $ in the two components \cite{we}. Numerical
solutions for them are produced, severally, by the following inputs:%
\begin{equation}
\phi _{+}^{(0)}=A_{1}\exp \left( -\alpha _{1}r^{2}\right) ,\phi
_{-}^{(0)}=A_{2}r\exp \left( i\theta -\alpha _{2}r^{2}\right) ,
\label{ansatz}
\end{equation}%
\begin{equation}
\overline{\phi _{+}^{(0)}}=-A_{2}r\exp \left( -i\theta -\alpha
_{2}r^{2}\right) ,\overline{\phi _{-}^{(0)}}=A_{1}\exp \left( -\alpha
_{1}r^{2}\right) ,  \label{ansatz2}
\end{equation}%
which are written in polar coordinates $\left( r,\theta \right) $, with real
$A_{1,2}$ and $\alpha _{1,2}>0$. The creation of MMs and CSs is initiated by
a combination of inputs (\ref{ansatz}) and (\ref{ansatz2}) with the opposite
chiralities,%
\begin{equation}
\phi _{\pm }=M\phi _{\pm }^{(0)}+\sqrt{1-M^{2}}\cdot \overline{\phi _{\pm
}^{(0)}},  \label{M}
\end{equation}%
where the weight factor takes values $-1\leq M\leq 1$. The choice of $|M|=1/%
\sqrt{2}$ in Eq. (\ref{M}) corresponds to MMs, while $|M|\neq 0,1/\sqrt{2}$
corresponds to CSs (in the case of $\gamma =1$). In our double-layer system,
the symmetric and asymmetric solutions can be generated by setting the
initial norms in the layers as $N_{\phi }=N_{\psi }$ and $N_{\phi }\neq
N_{\psi }$, respectively.

\section{Symmetric and asymmetric solitons in the double-layer system}

\subsection{Semi-vortices}

\begin{figure}[tbp]
\centering{\label{fig1a} \includegraphics[scale=0.14]{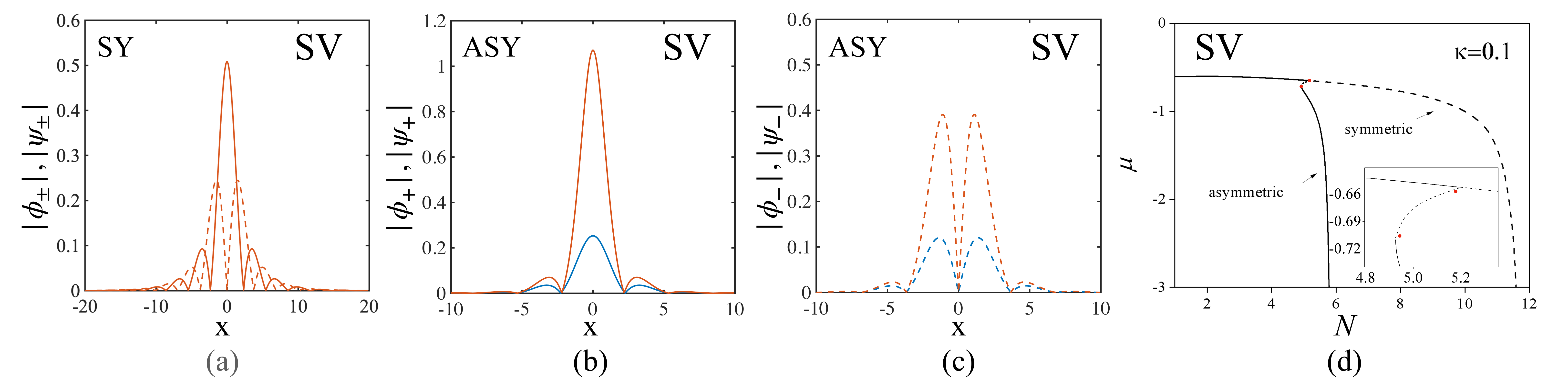}}
\caption{(Color online) (a) Cross-section profiles, drawn along $y=0$, of a
stable symmetric semi-vortex (SV) with $\protect\gamma =0$ in Eqs. (\protect
\ref{phipsi}), total norm $N=5$, and inter-layer coupling constant $\protect%
\kappa =0.1$. The solid and dashed lines represent, respectively, $%
\left\vert \protect\phi _{+}\right\vert =\left\vert \protect\psi %
_{+}\right\vert $ and $\left\vert \protect\phi _{-}\right\vert =\left\vert
\protect\psi _{-}\right\vert $. Here and in other figures, labels SY and ASY
pertain, respectively, to symmetric and asymmetric solitons in the
double-layer system. Panels (b) and (c) separately display the profiles
(solid and dashed, for $\protect\phi _{+},\protect\psi _{+}$ and $\protect%
\phi _{-},\protect\psi _{-}$, respectively) for a stable asymmetric SV, with
the same values of $N$ and $\protect\kappa $. In panels (b) and (c), blue
and orange curves represent, respectively, $\left\vert \protect\phi _{\pm
}\right\vert $ and $\left\vert \protect\psi _{\pm }\right\vert $ components,
with the broken symmetry between the coupled layers ($\left\vert \protect%
\phi _{\pm }\right\vert \neq \left\vert \protect\psi _{\pm }\right\vert $).
These symmetric and asymmetric SVs belong to the bistability region, see Eq.
(\protect\ref{bi}). (d) The chemical potential, $\protect\mu $, vs. the
total norm, $N$, for the families of the symmetric and asymmetric SV
solitons. Here, the solid and dashed curves designate stable and unstable
solutions, respectively. Red dots denote the symmetry-breaking bifurcation
point, at $N=N_{\mathrm{cr}}$, and the turning point, at $N=N_{\mathrm{turn}%
} $, of the $\protect\mu (N)$ curve for the asymmetric family, see Eq. (%
\protect\ref{bi}). The inset provides a zoom of the area around the
symmetry-breaking bifurcation.}
\label{SVs}
\end{figure}

\begin{figure}[tbp]
\centering{\label{fig2a} \includegraphics[scale=0.12]{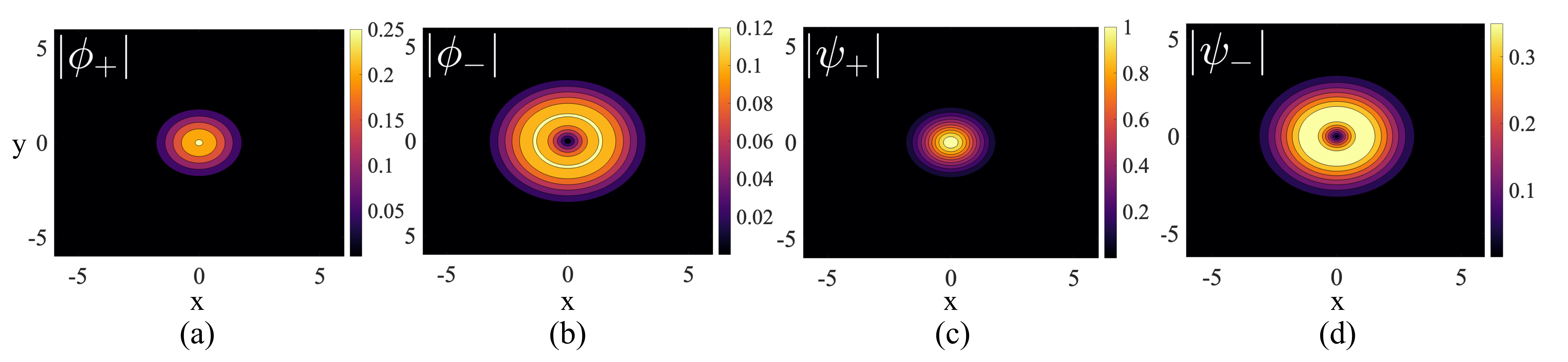}}
\caption{(Color online) (a) Top-view profiles of the absolute value of the
zero-vorticity (a,c) and vortical (b,d) components of the same stable
asymmetric SV which is represented by its cross-section profiles in Figs.
\protect\ref{SVs}(b,c).}
\label{2DASVs}
\end{figure}

\begin{figure}[tbp]
\centering{\label{fig3a} \includegraphics[scale=0.23]{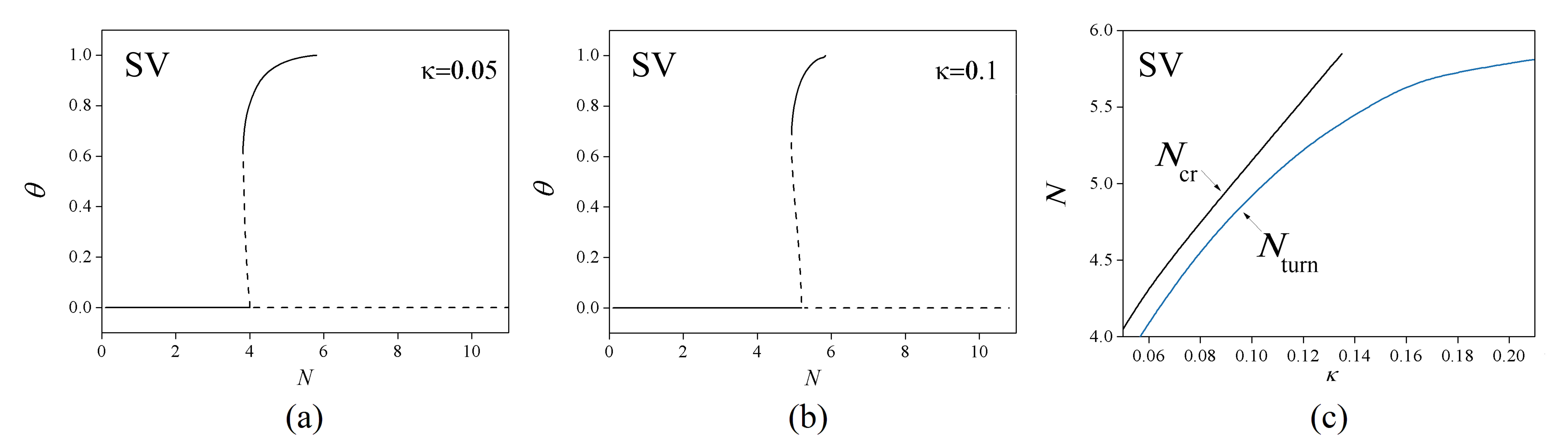}}
\caption{(Color online) (a,b) The dependence of the asymmetry parameter (%
\protect\ref{ratio}) on the total norm (\protect\ref{Norm}) for the
two-layer 2D solitons of the SV type, with $\protect\gamma =0$ in Eqs. (%
\protect\ref{phipsi}) and values of the inter-layer coupling $\protect\kappa %
=0.05$ (a) or $\protect\kappa =0.1$ (b). Solid and dashed curves designate
stable and unstable families of the solutions, respectively. (c) The value
of the norm at the turning point, $N_{\mathrm{turn}}$, at which the
asymmetric SV states are created by the subcritical bifurcation, and the
critical norm, $N_{\mathrm{cr}}$, at which the symmetric SV solution becomes
unstable, vs. $\protect\kappa $. The system is bistable, featuring the
coexistence of stable symmetric and asymmetric states, in the region of $N_{%
\mathrm{turn}}<N<$ $N_{\mathrm{cr}}$, see Eq. (\protect\ref{bi}).}
\label{SV_theta_kappa}
\end{figure}

Basic numerical findings for SV solitons are displayed in Figs. \ref{SVs}
and \ref{SV_theta_kappa} [recall the numerical results are produced setting $%
\lambda =1$ in Eqs. (\ref{phipsi})]. First, typical shapes of stable SVs,
which are symmetric and asymmetric with respect to the two layers, are
displayed, by means of their 1D cross sections, in Figs. \ref{SVs}(a) and
(b,c), respectively. 2D profiles of the same asymmetric SV which is
presented in Figs. \ref{SVs}(b,c) are displayed in Fig. \ref{2DASVs}. They
feature a maximum at the center ($x=y=0$) in the $\left( \phi _{+},\psi
_{+}\right) $ components, and zero at the same point in the $\left( \phi
_{-},\psi _{-}\right) $ ones, in accordance with the basic structure of the
SV: it contains a zero-vorticity mode in $\left( \phi _{+},\psi _{+}\right) $%
, and a vortical one in $\left( \phi _{-},\psi _{-}\right) $.

Families of stable and unstable symmetric and asymmetric SV modes are
represented by curves showing the modes' chemical potential. $\mu $, vs.
their total norm (\ref{Norm}) in Fig. \ref{SVs}(d). Note that the short
segment of the asymmetric family, between the SSB point and the turning one,
features $d\mu /dN>0$, while its extension, in the form of the long segment
below the turning point, has $d\mu /dN<0$. In agreement with the
Vakhitov-Kolokolov (VK) criterion, which is a well-known necessary stability
condition \cite{VK,Sulems,Z and K,Fibich}, these segments represent,
respectively, unstable and stable states. The entire branch of the symmetric
SV solitons satisfies the VK criterion, but it is unstable against SSB above
the critical point (the latter instability cannot be captured by the VK
criterion). As is typical for subcritical bifurcations \cite{Iooss}, the
interval of values of the norm,%
\begin{equation}
N_{\mathrm{turn}}<N<N_{\mathrm{cr}},  \label{bi}
\end{equation}%
between the turning and critical (bifurcation) points maintains bistability,
as both the symmetric and asymmetric states are stable in it.

The branch of the symmetric solitons naturally terminates at the point of
\begin{equation}
N=2N_{\mathrm{TS}},  \label{N=2NTS}
\end{equation}%
with norm $N_{\mathrm{TS}}$ in each layer. On the other hand, the branch
representing the asymmetric solitons terminates when the asymmetry parameter
(\ref{ratio}) attains the limit, $\theta =1$, i.e., one layer becomes empty
and, accordingly, the respective total norm is $N=N_{\mathrm{TS}}$.
Comparison of Fig. \ref{SVs}(d) with the known value $N_{\mathrm{TS}}\approx
5.85$ [see Eq. (\ref{NTS})] corroborates these expectations.

The SSB bifurcation in the double-layer systems is characterized by the
dependence of the asymmetry parameter (\ref{ratio}) on the full norm (\ref%
{Norm}) \cite{Pare}-\cite{Warsaw}. It is displayed in Fig. \ref%
{SV_theta_kappa} for the two-layer solitons of the SV type with two
different values of the inter-layer coupling constant, \textit{viz}., $%
\kappa =0.05$ (a) and $\kappa =0.1$ (b). It is obvious that the bifurcation
is of the subcritical type, featuring the turning point on the $\theta (N)$
curves, and narrow bistability regions identified as per Eq. (\ref{bi}). The
numerical results produce similar $\theta (N)$ curves at other values of $%
\kappa $ -- in particular, the bifurcation never becomes supercritical, in
which the curve would not go backward, and no bistability would be observed.
Overall, the subcritical SSB bifurcation is described by dependences $N_{%
\mathrm{turn}}(\kappa )$ and $N_{\mathrm{cr}}(\kappa )$, which are displayed
in Fig. \ref{SV_theta_kappa}(c). The entire bistability region is located
between these two curves. All the curves shown in Fig. \ref{SV_theta_kappa}
for the asymmetric families terminate at the above-mentioned point, $N=N_{%
\mathrm{TS}}\approx 5.85$, at which the critical collapse sets in. The
collapse does not admit the existence of stationary solitons at $N>N_{%
\mathrm{TS}}$.

\begin{figure}[tbp]
\centering{\label{fig4a} \includegraphics[scale=0.5]{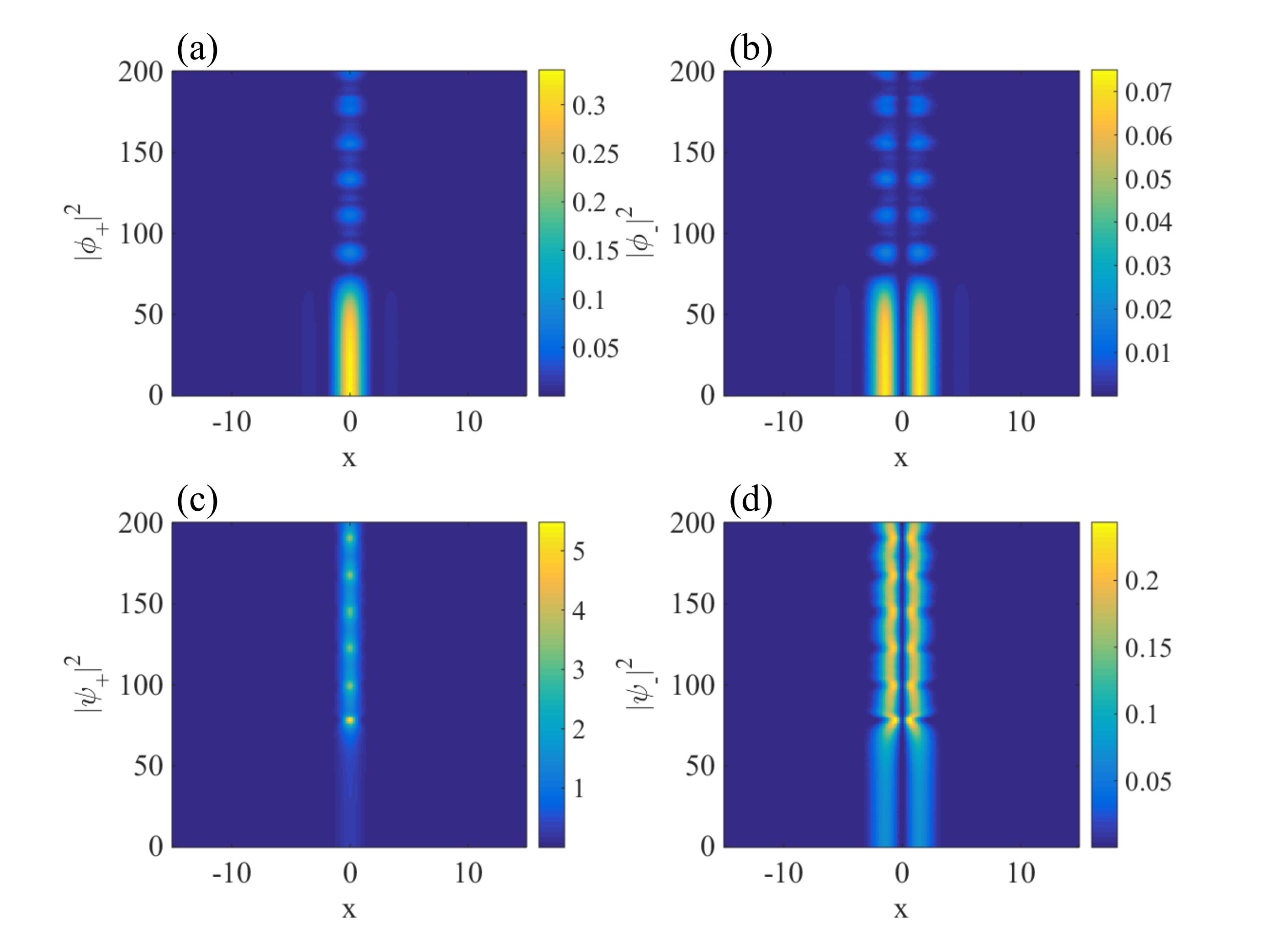}}
\caption{(Color online) The numerically simulated evolution of an unstable
symmetric SV (shown by density cross-sections of its components along $y=0$)
with $\protect\kappa =0.1$ and $N=5.6$. The evolution exhibits the onset of
the SSB\ (spontaneous symmetry breaking) with concomitant Josephson
oscillations. Note that these values of the parameters belong to interval (%
\protect\ref{NNN}). Note also that widely different scales are used for
plotting the component densities in the two layers, $\left\vert \protect\phi %
_{\pm }\right\vert ^{2}$ and $\left\vert \protect\psi _{\pm }\right\vert
^{2} $, because the SSB makes their amplitudes strongly different.}
\label{SV_DR1}
\end{figure}

In direct simulations, unstable symmetric solitons of the SV type with the
total norm falling in interval
\begin{equation}
N_{\mathrm{cr}}<N<N_{\mathrm{TS}}  \label{NNN}
\end{equation}%
(hence the collapse cannot commence in the system) demonstrate dynamical
symmetry breaking, spontaneously transforming into oscillatory states
(breathers) which keep the original structure of the SV soliton, i.e., the
zero-vorticity and vorticity-$1$ shapes of their larger and smaller
components, while the overall amplitude of the SV mode spontaneously becomes
higher in one layer than in the other. A typical example of such a breather
with the spontaneously broken inter-layer symmetry is displayed in Fig. \ref%
{SV_DR1}. This dynamical regime may be considered as Josephson oscillations
in the bosonic junction, cf. Refs. \cite{Milburn,Smerzi,Markus}. The
oscillations proceed out-of-phase in the tunnel-coupled layers, i.e., a
maximum amplitude in one layer coincides, in time, with a minimum in the
other one. Naturally, the oscillation frequencies grow with the increase of
the solitons' norm. Detailed results of the simulations demonstrate that the
spontaneously established oscillations are accompanied by generation of
dispersive waves (\textquotedblleft radiation") with very small amplitudes.
The radiation is virtually invisible in Fig. \ref{SV_DR1}, and remained
invisible as long as the simulations were running.

\begin{figure}[tbp]
\centering{\label{fig5a} \includegraphics[scale=0.5]{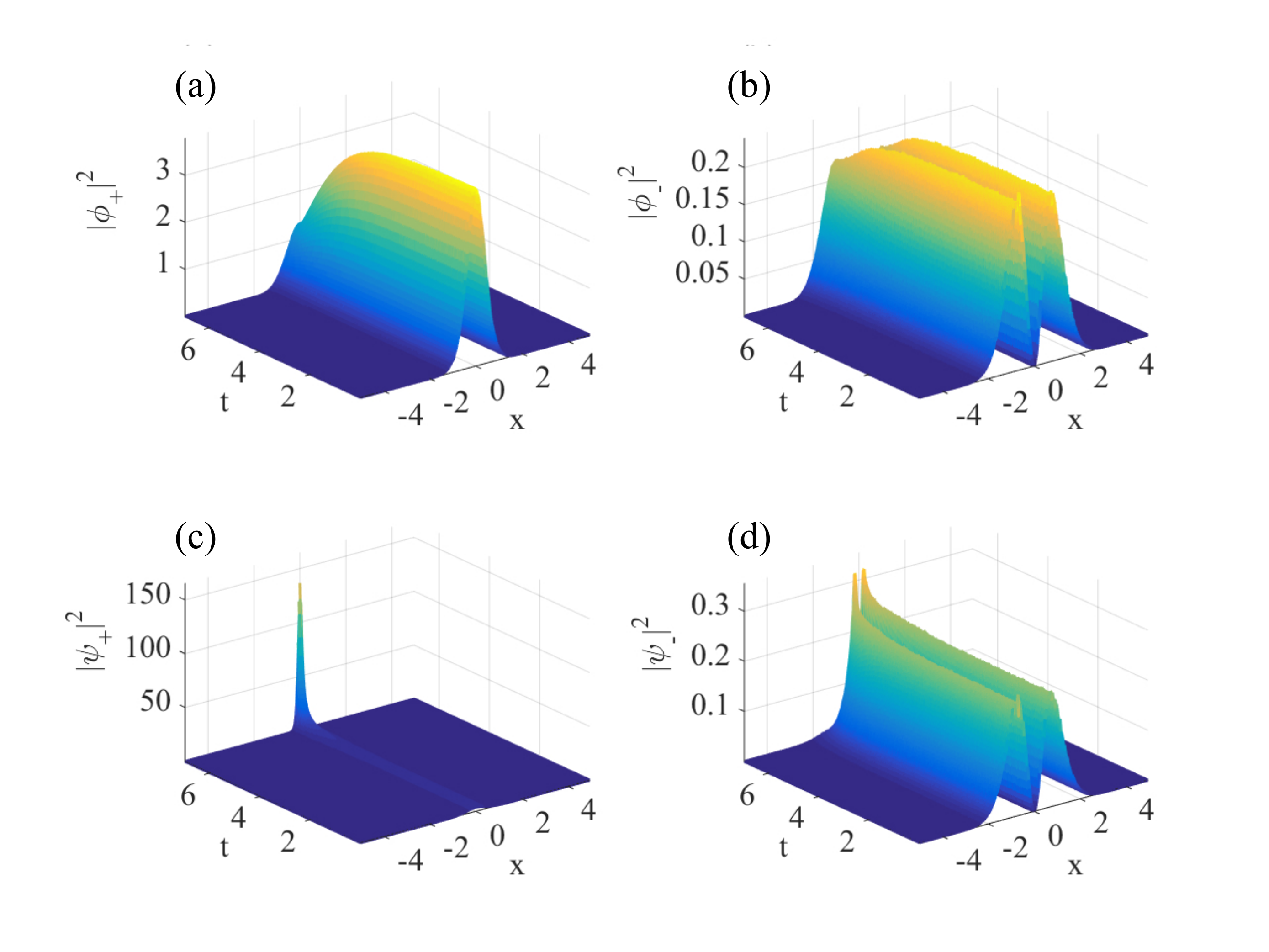}}
\caption{(Color online) The evolution of an unstable symmetric SV (shown by
means of the density cross-sections along $y=0$) with $\protect\kappa =0.1$
and $N=11$. The latter value falls in interval (\protect\ref{NTS<N<2NTS}),
where the collapse may take place in a single layer, which indeed happens
here. To display the evolution of the intensities in the two layers, very
different scales are used in the plots for $|\protect\phi _{\pm }|^{2}$ and $%
|\protect\psi _{\pm }|^{2}$.}
\label{SV_DR2}
\end{figure}

The outcome of the development of the spontaneously initiated dynamical
symmetry breaking is different in the interval of%
\begin{equation}
N_{\mathrm{TS}}<N<2N_{\mathrm{TS}}.  \label{NTS<N<2NTS}
\end{equation}%
Instead of the oscillations, the growth of the norm in one layer allows it
to reach the collapse threshold, which indeed leads to the blowup of the
fields, see Fig. \ref{SV_DR2}.

\begin{figure}[tbp]
\centering{\label{fig6a} \includegraphics[scale=0.5]{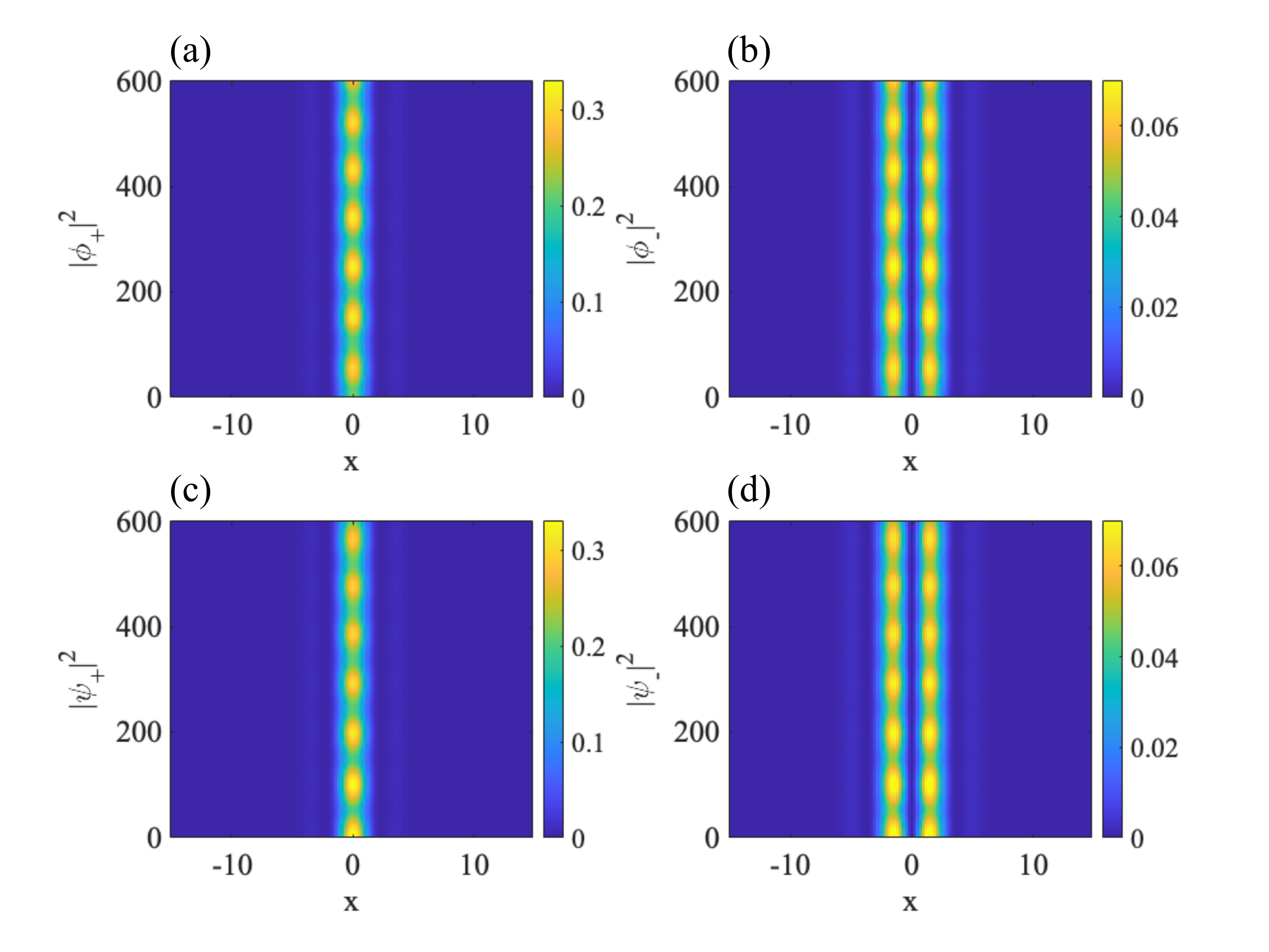}}
\caption{(Color online) The numerically simulated evolution of an unstable
asymmetric SV (shown by the density cross-section along $y=0$) with $\protect%
\kappa =0.1$ and $N=5.12$. Note that this value belongs to interval (\protect
\ref{NNN}). The evolution exhibits the spontaneous transformation of the
unstable SV into a breather oscillating around the stable symmetric SV.}
\label{ASV_DR1}
\end{figure}

It is relevant to stress the SSB for SVs takes place in a finite interval of
the values of the coupling constant,
\begin{equation}
0<\kappa <\kappa _{\max }^{\mathrm{(SV)}}\approx 0.135,  \label{kappa-max}
\end{equation}%
with $\kappa _{\max }^{\mathrm{(SV)}}$ determined by the condition that the
SSB occurs at $N=N_{\mathrm{TS}}$, i.e., by equation
\begin{equation}
N_{\mathrm{cr}}\left( \kappa _{\max }^{\mathrm{(SV)}}\right) =N_{\mathrm{TS}%
},  \label{NNTS}
\end{equation}%
see Fig. \ref{SV_theta_kappa}(c). At $\kappa >\kappa _{\max }^{\mathrm{(SV)}%
} $, the critical collapse occurs prior to the expected onset of the SSB.

As concerns unstable asymmetric states existing in interval (\ref{bi}),
which are shown by dashed segments of the respective curves $\mu (N)$ in
Fig. \ref{SVs}(d) and $\theta (N)$ in Figs. \ref{SV_theta_kappa}(a,b),
direct simulations demonstrate that, depending on values of the parameters
and small perturbations which initiate the growth of the instability, they
evolve into breathers which oscillate around either the stable symmetric
state or stable asymmetric one, coexisting with its symmetric counterpart in
the bistability region (\ref{bi}). A typical example of the oscillation
instability is shown in Fig. \ref{ASV_DR1}.

\subsection{Mixed modes}

\begin{figure}[tbp]
\centering{\label{fig7a} \includegraphics[scale=0.14]{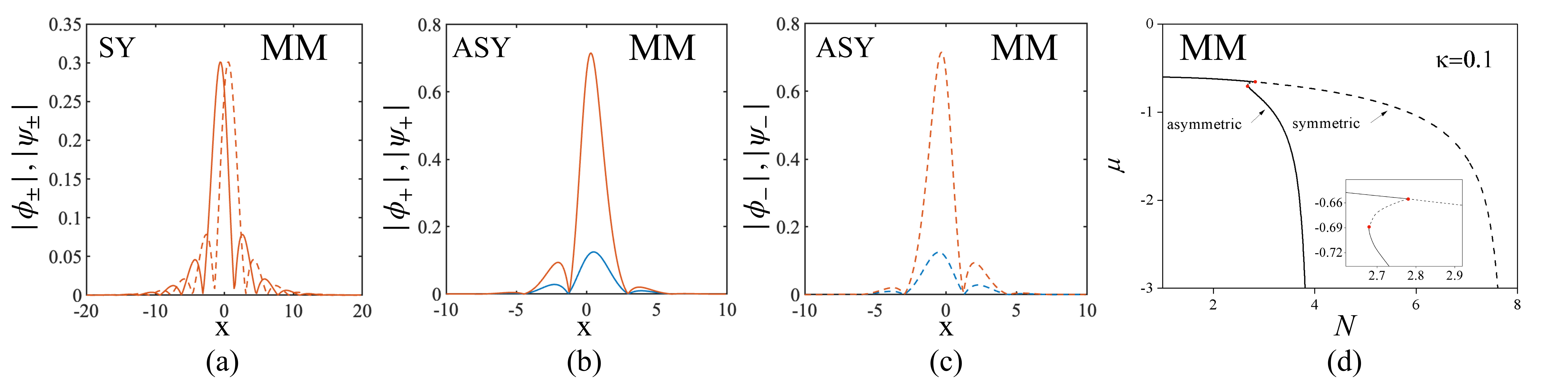}}
\caption{(Color online) (a) Cross-section profiles, along axis $y=0$, of a
stable symmetric MM\ (mixed mode) with $\protect\gamma =2$ in Eqs. (\protect
\ref{phipsi}), total norm $N=2.7$, and the inter-layer coupling constant $%
\protect\kappa =0.1$. Solid and dashed lines designate, respectively, $%
\left\vert \protect\phi _{+}\right\vert =\left\vert \protect\psi %
_{+}\right\vert $ and $\left\vert \protect\phi _{-}\right\vert =\left\vert
\protect\psi _{-}\right\vert $. (b,c) The same profiles (solid and dashed
for $\protect\phi _{+},\protect\psi _{+}$ and $\protect\phi _{-},\protect%
\psi _{-}$, respectively) for a stable asymmetric MM with $N=3$ and $\protect%
\kappa =0.1$. Here blue and orange curves represent, respectively, the $%
\protect\phi _{\pm }$ and $\protect\psi _{\pm }$ components. (d) $N(\protect%
\mu )$ dependences for the families of symmetric and asymmetric two-layer
states of the MM type. In this panel, solid and dashed curves curves stand
for stable and unstable solutions, respectively. Dots denote the
symmetry-breaking bifurcation point, and the turning point of the $\protect%
\mu (N)$ curve for the asymmetric family. The inset provides a zoom of the
area around the bifurcation, cf. Fig. \protect\ref{SVs} for the SV states.}
\label{MMs}
\end{figure}

\begin{figure}[tbp]
\centering{\label{fig8a} \includegraphics[scale=0.12]{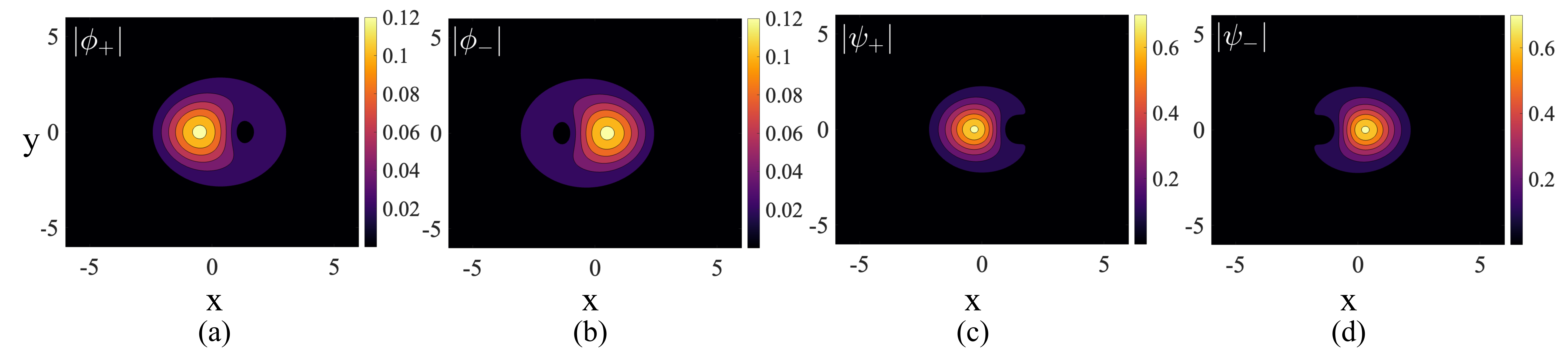}}
\caption{(Color online) Top-view profiles of the components of the
asymmetric MM which is presented by its cross-section profiles in Figs.
\protect\ref{MMs} (b,c), cf. Fig. \protect\ref{2DASVs} for the asymmetric
SV. }
\label{2DAMM}
\end{figure}

Typical examples of amplitude profiles in stable symmetric and asymmetric
MMs are plotted, by means of the cross sections along $y=0$, in Figs. \ref%
{MMs}(a) and (b,c), respectively. 2D profiles of the same asymmetric MM
which is presented in Figs. \ref{MMs}(b,c) are displayed in Fig. \ref{2DAMM}%
. Note that, while the system spontaneously breaks the symmetry between the
tunnel-coupled layers, the solutions keep the cross symmetries of Eqs. (\ref%
{PhiPsi}) with respect to the following substitution: $\left\{ \Phi ,\Psi
\right\} _{+}\longleftrightarrow \left\{ \Phi ,\Psi \right\}
_{-},x\rightarrow -x$, and $\left\{ \Phi ,\Psi \right\}
_{+}\longleftrightarrow -\left\{ \Phi ,\Psi \right\} _{-},y\rightarrow -y$.

Dependences $N(\mu )$ for the two-layer MM states are displayed in Fig. \ref%
{MMs}(d). Note that the VK criterion, if applied to these dependences,
produces essentially the same conclusions as reported above for the SV
families shown in Fig. \ref{SVs}. The straightforward consideration which is
also similar to that presented above for the SVs demonstrates that the
asymmetric MM states exist in the interval of values of the total norm given
by Eq. (\ref{MM-Townes}), which is bounded by the onset of the collapse in
the single layer, while the symmetric MM solitons exist in the interval
which is twice as broad, \textit{viz}., $0<N<4N_{\mathrm{TS}}/(\gamma +1)$,
cf. Eq. (\ref{N=2NTS}).

\begin{figure}[tbp]
\centering{\label{fig9a} \includegraphics[scale=0.23]{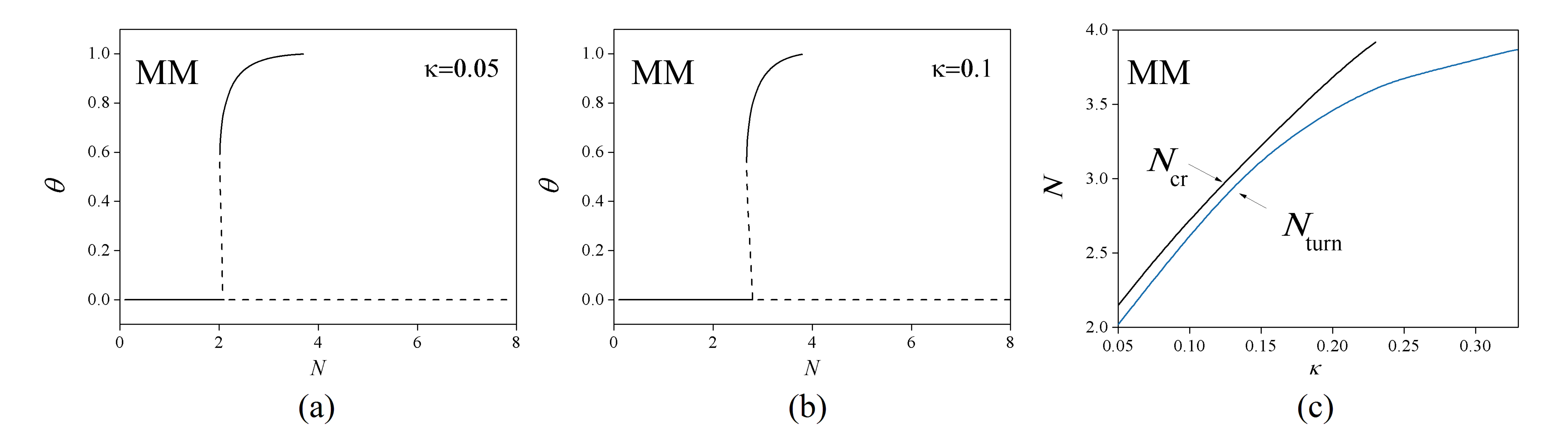}}
\caption{(Color online) The subcritical SSB bifurcation of 2D two-layer MMs,
for $\protect\gamma =2$ in Eqs. (\protect\ref{phipsi}) and $\protect\kappa %
=0.05$ (a) or $\protect\kappa =0.1$ (b), is illustrated by dependences of
asymmetry (\protect\ref{ratio}) on the total norm (\protect\ref{Norm}), cf.
Figs. \protect\ref{SV_theta_kappa}(a,b). Solid and dashed curves designate
stable and unstable solutions, respectively. (c) The value of the norm at
the turning point, $N_{\mathrm{turn}}$, at which the asymmetric MM states
are created by the subcritical bifurcation, and the critical norm, $N_{%
\mathrm{cr}}$, at which the symmetric MM solution becomes unstable, vs. $%
\protect\kappa $. The system is bistable, featuring the coexistence of
stable symmetric and asymmetric states, in the region of $N_{\mathrm{turn}%
}<N<$ $N_{\mathrm{cr}}$, cf. Fig. \protect\ref{SV_theta_kappa}(c) for the
two-layer SV solitons.}
\label{MM_theta_kappa}
\end{figure}

The SSB bifurcation for the MM states is also of the subcritical type, as
shown in Figs. \ref{MM_theta_kappa}(a,b). Similar to the case of the SVs,
the bifurcation for the MMs is characterized by dependences $N_{\mathrm{turn}%
}(\kappa )$\ and $N_{\mathrm{cr}}(\kappa )$, with the bistability region
bounded by these curves, as shown in Fig. \ref{MM_theta_kappa}(c).

Figure \ref{MM_theta_kappa}(c) demonstrates that the SSB for MMs takes place
in a finite interval of the values of the inter-layer coupling constant,
\begin{equation}
0<\kappa <\kappa _{\max }^{\mathrm{(MM)}}\approx 0.23  \label{kappa-max-MM}
\end{equation}%
[cf. Eq. (\ref{kappa-max}) for the SV states], with $\kappa _{\max }^{%
\mathrm{(MM)}}$ determined by the condition
\begin{equation}
N_{\mathrm{cr}}\left( \kappa _{\max }^{\mathrm{(MM)}}\right) =2N_{\mathrm{TS}%
}/\left( 1+\gamma \right) ,  \label{NNTSMM}
\end{equation}%
cf. Eq. (\ref{NNTS}). At $\kappa >\kappa _{\max }^{\mathrm{(MM)}}$, the SSB
does not take place, as the critical collapse occurs earlier. Accordingly,
all curves for asymmetric MMs in Fig. \ref{MM_theta_kappa} terminate at $%
N=(2/3)N_{\mathrm{TS}}$, which corresponds to the value in Eq. (\ref{NNTSMM}%
) at $\gamma =2$.

It is worthy to note that the existence region of the SSB for the MM
solitons, as given by Eq. (\ref{kappa-max-MM}) is essentially broader than
the similar region for the SV states, given by Eq. (\ref{kappa-max}).

\begin{figure}[tbp]
\centering{\label{fig10a} \includegraphics[scale=0.11]{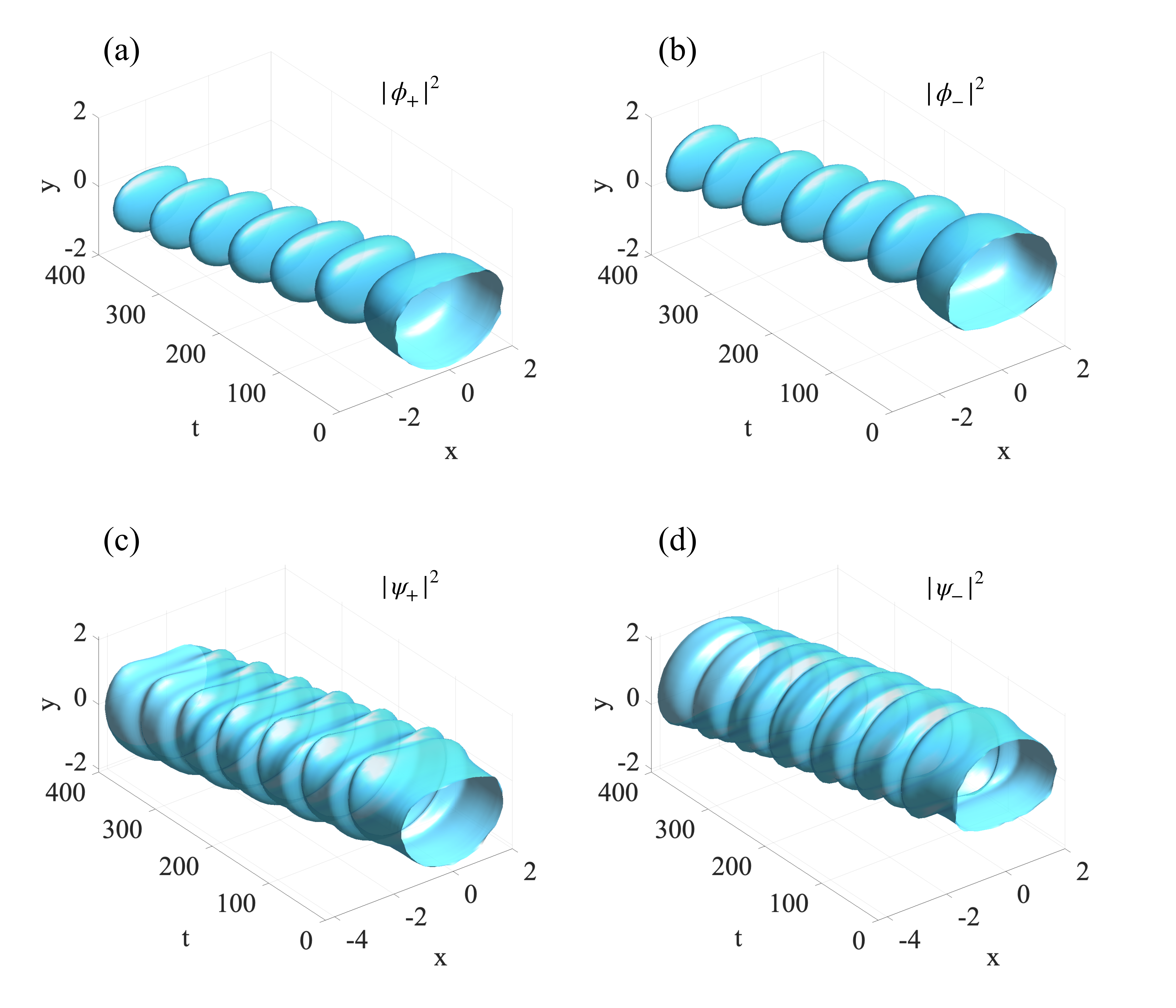}}
\caption{(Color online) The evolution of an unstable symmetric MM, with $%
\protect\gamma =2$, $\protect\kappa =0.1$, and $N=3$, which belongs to
interval (\protect\ref{NMMunst}). The spontaneously established intrinsic
oscillations and slow drift of the two-layer 2D soliton are displayed by
means surface plots $\left\vert \protect\phi _{\pm }(x,y)\right\vert
^{2}=\left\vert \protect\psi _{\pm }\left( x,y\right) \right\vert ^{2}=0.03$%
. }
\label{MM_osci}
\end{figure}

Direct simulations, displayed in Fig. \ref{MM_osci}, demonstrates that the
symmetry-breaking instability of symmetric MM solitons, which exist in the
interval of
\begin{equation}
N_{\mathrm{cr}}<N<2N_{\mathrm{TS}}/\left( 1+\gamma \right) ,  \label{NMMunst}
\end{equation}%
gives rise to breathers, accompanied by the emission of very weak dispersive
waves (practically invisible in Fig. \ref{MM_osci}). Their behavior is very
different from that of the breathers generated by the instability if the
symmetric states of the SV type, cf. Fig. \ref{SV_DR1}. Indeed, along with
the intrinsic oscillations, the breathers of the MM type feature slow
\textit{spontaneous drift}, which is clearly seen in Fig. \ref{MM_osci}. The
trend of unstable MMs to develop spontaneous drift is also known in other
cases \cite{Ben-Li}.

In addition to interval (\ref{NMMunst}), unstable symmetric MMs exist also
in the range of%
\begin{equation}
2N_{\mathrm{TS}}/\left( 1+\gamma \right) <N<4N_{\mathrm{TS}}/\left( 1+\gamma
\right) ,  \label{NTS<N<2NTS-MM}
\end{equation}%
where the collapse may commence in a single layer, cf. a similar interval (%
\ref{NTS<N<2NTS}) for the SV solitons. Accordingly, in this case the
development of the SSB leads to the collapse, as shown in Fig. \ref%
{MM_collapse}, similar to what happens with the unstable symmetric SV
soliton in Fig. \ref{SV_DR2}.

\begin{figure}[tbp]
\centering{\label{fig11a} \includegraphics[scale=0.5]{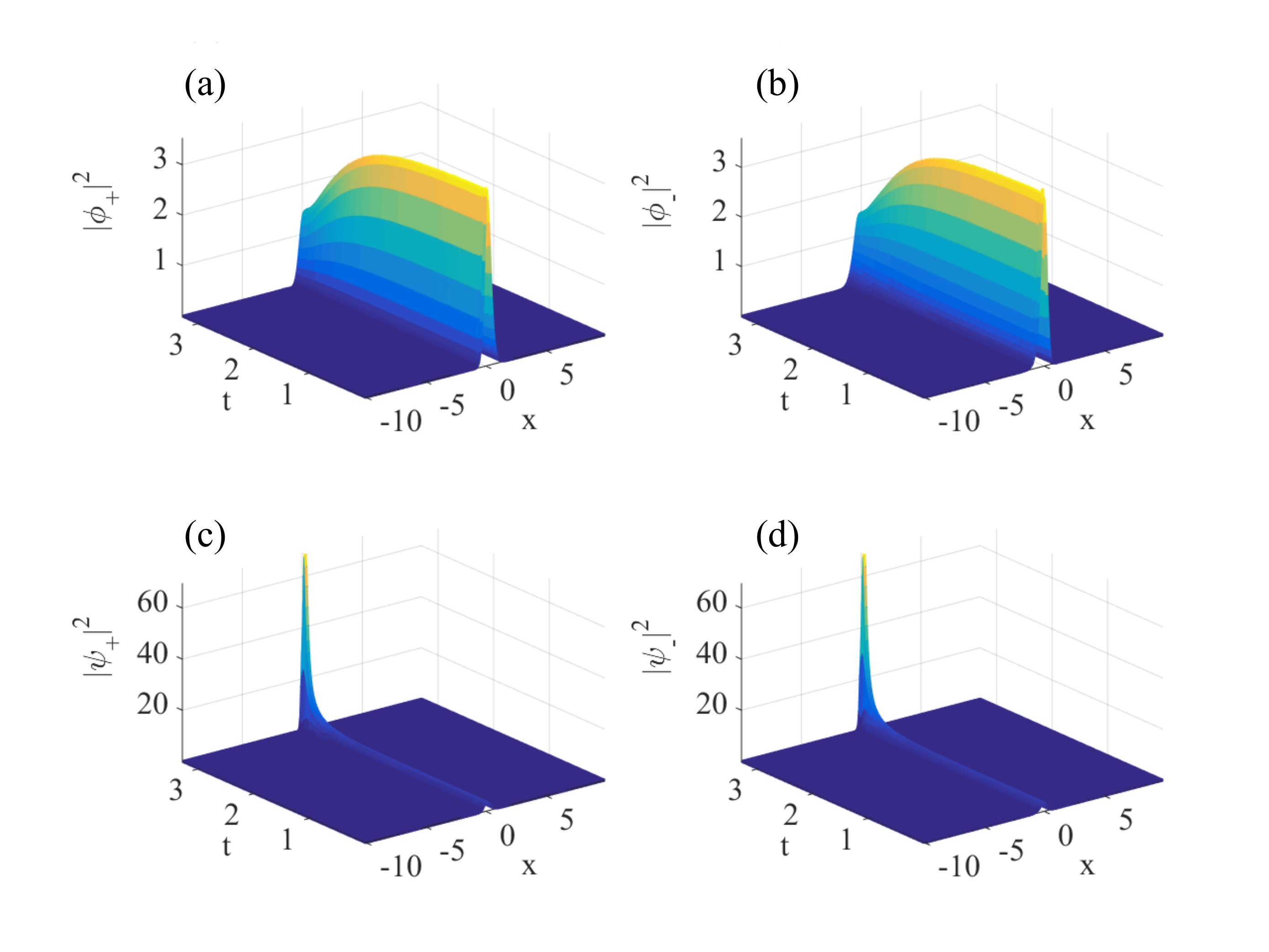}}
\caption{(Color online) The evolution of an unstable symmetric MM (shown by
its density cross-section along $y=0$) with $\protect\gamma =2$, $\protect%
\kappa =0.1$, and $N=7.6$, which belongs to interval (\protect\ref%
{NTS<N<2NTS-MM}). In this case, the onset of the SSB quickly leads to the
collapse in one layer. }
\label{MM_collapse}
\end{figure}

\subsection{Composite solitons}

\begin{figure}[tbp]
\centering{\label{fig12a} \includegraphics[scale=0.6]{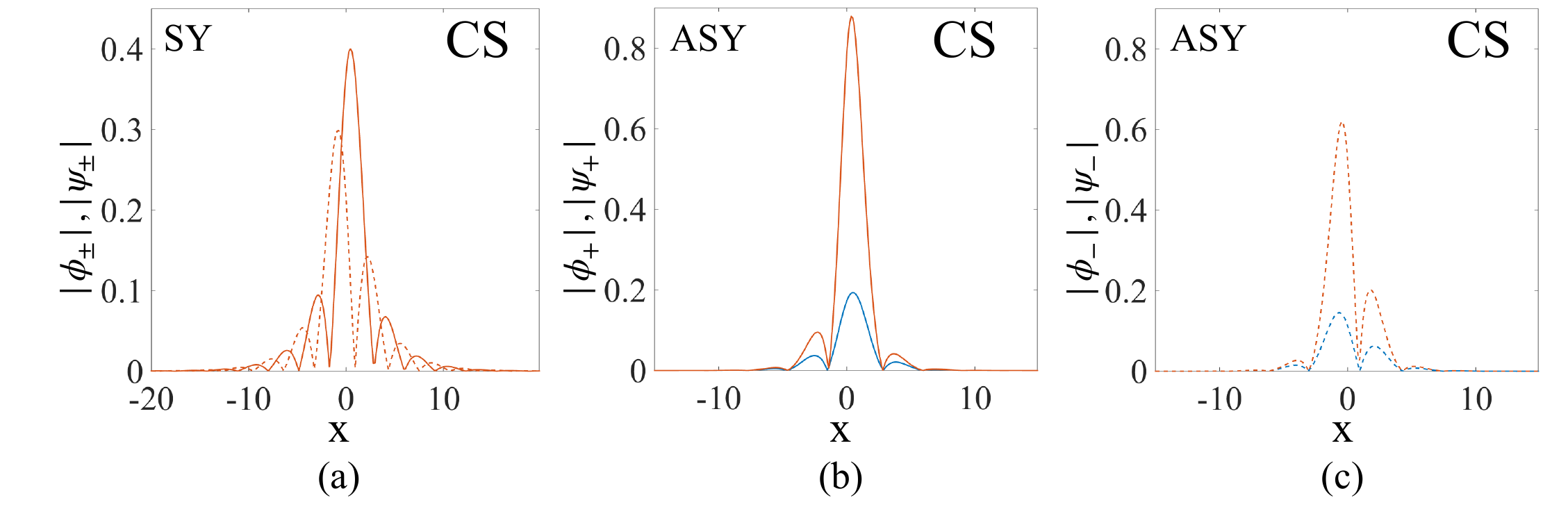}}
\caption{(Color online) (a) Cross-section profiles (along axis $y=0$) of a
stable symmetric CS (composite soliton) with $\protect\gamma =1$ in Eqs. (%
\protect\ref{phipsi}), $N=3.6$ and $\protect\kappa =0.1$. Values of the
angular momentum per particle [defined as per Eq. (\protect\ref{L})] for the
components are $(L_{\protect\phi _{+}},L_{\protect\phi _{-}})=(-0.23,0.75)$.
The solid and dashed lines designate, respectively, $\left\vert \protect\phi %
_{+}\right\vert =\left\vert \protect\psi _{+}\right\vert $ and $\left\vert
\protect\phi _{-}\right\vert =\left\vert \protect\psi _{-}\right\vert $.
(b,c) The same profiles for a stable asymmetric CS with $\protect\gamma =1$,
$N=4$, and $\protect\kappa =0.1$ (solid and dashed for $\protect\phi _{+},%
\protect\psi _{+}$ and $\protect\phi _{-},\protect\psi _{-}$, respectively).
In this case, orange and blue curves represent the $\protect\phi _{\pm }$
and $\protect\psi _{\pm }$ components, respectively. Angular momenta per
particle for the four components are $(L_{\protect\phi _{+}},L_{\protect\phi %
_{-}},L_{\protect\psi _{+}},L_{\protect\psi _{-}})=$ $%
(-0.26,0.72,-0.22,0.67).$}
\label{CS}
\end{figure}

\begin{figure}[tbp]
\centering{\label{fig13a} \includegraphics[scale=0.12]{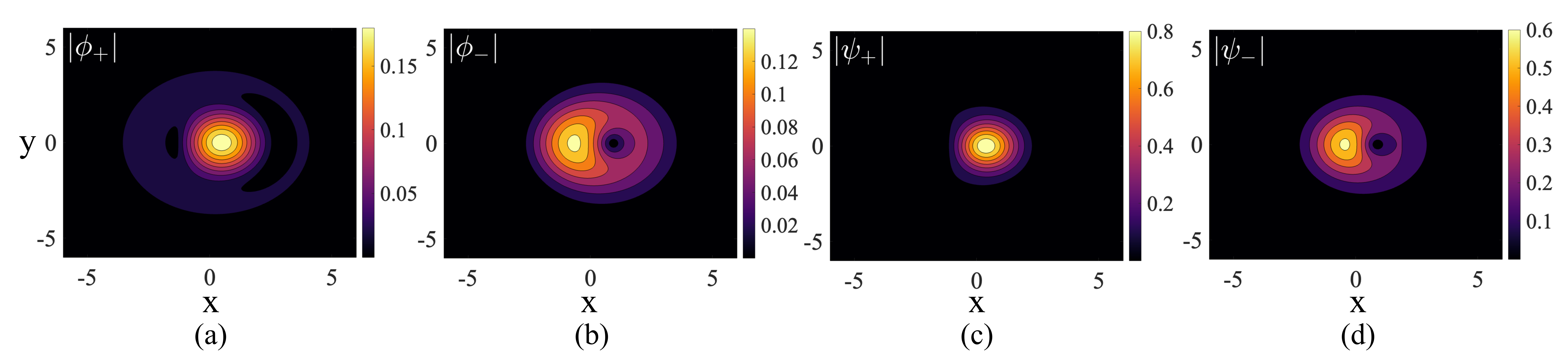}}
\caption{(Color online) (a) Top-view profiles of components of the
asymmetric CS which is presented by its cross-section profiles in Figs.
\protect\ref{CS}(b,c), cf. Fig. \protect\ref{2DAMM} for the asymmetric MM.}
\label{2DACS}
\end{figure}

As well as its single-layer counterpart \cite{Ben-Li}, the present system
with $\gamma =1$ in Eqs. (\ref{phipsi}) supports a continuous family of CS
states connecting the SV and MM solutions with equal values of $N$ and $\mu $%
. Typical examples of stable symmetric and asymmetric CSs are displayed in
Figs. \ref{CS}(a) and (b,c), respectively, with the top views of the
component distribution in the latter one plotted in Fig. \ref{2DACS}.

As mentioned above, the CS family is a degenerate one, in the sense that all
solutions belonging to it share a single value of the total norm, as well as
a single value of the chemical potential. Within the family, the solutions
differ by values of the angular momentum per particle in each component,
\begin{equation}
L_{(\phi ,\psi )\pm }=N_{\pm }^{-1}\int \int \left( \phi ,\psi \right) _{\pm
}^{\ast }\hat{\mathcal{L}_{z}}\left( \phi ,\psi \right) _{\pm }dxdy,
\label{L}
\end{equation}%
with $\hat{\mathcal{L}_{z}}=-i(x\partial /\partial y-y\partial /\partial
x)\equiv -i\partial /\partial \eta $, where $\eta $ is the angular
coordinate in the $\left( x,y\right) $ plane.

Taking into regard that $2/(1+\gamma )=1$ for $\gamma =1$, Eqs. (\ref%
{MM-Townes}) and (\ref{N=2NTS}) predict that the symmetric and asymmetric CS
families exist in intervals $0<N<2N_{\mathrm{TS}}$ and $N<N_{\mathrm{TS}}$,
respectively. These predictions are corroborated by numerical simulations.
The numerical results for symmetric and asymmetric CS modes, including the $%
\mu (N)$ and $\theta (N)$ dependences, and conclusions concerning the
evolution of unstable states, are quite similar to those reported above for
the SV and MM states in Figs. \ref{SVs}(d)-\ref{SV_DR2} and (\ref{MMs}(d)-(%
\ref{MM_collapse}) (therefore the results for the CS family are not
displayed here). In particular, the SSB bifurcation for the two-layer CS
states is present in the interval of values of the inter-layer coupling $%
0<\kappa <\kappa _{\max }^{\mathrm{(CS)}}\approx 0.27$, cf. Figs. \ref%
{SV_theta_kappa}(c) and \ref{MM_theta_kappa}(c), and Eqs. (\ref{kappa-max})
and (\ref{kappa-max-MM}), for the SV and MM families, respectively.

\subsection{Moving solitons}

\begin{figure}[tbp]
\centering{\label{fig14a} \includegraphics[scale=0.11]{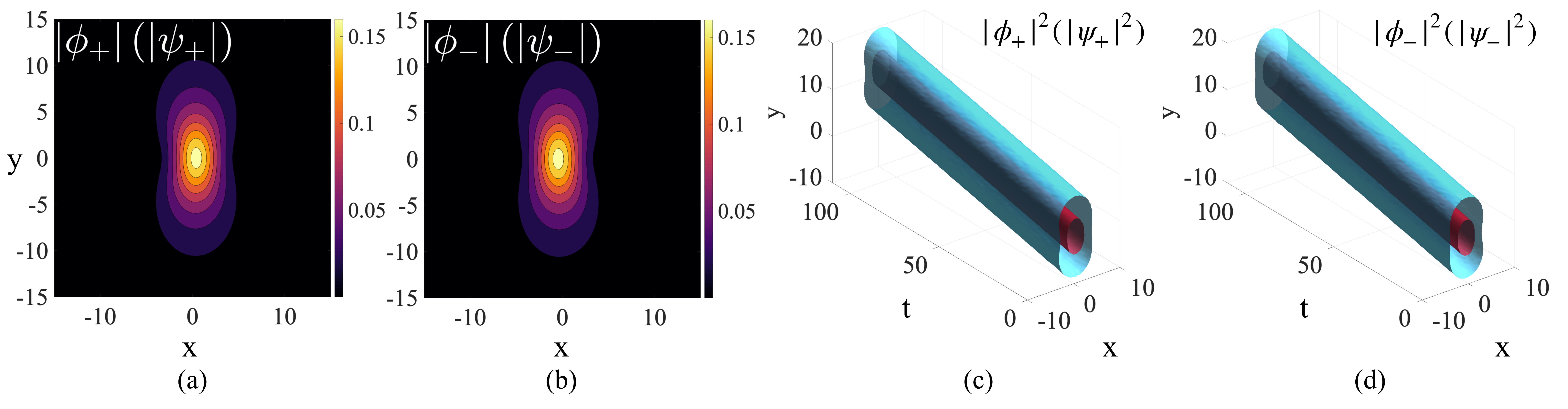}}
\caption{(Color online) Contour plots of $|\protect\phi _{+}(x,y)|=|\protect%
\psi _{+}(x,y)|$ (a) and $|\protect\phi _{-}(x,y)|=|\protect\psi _{-}(x,y)|$
(b) of a stable symmetric MM soliton, with $N=3$, and parameters $\protect%
\gamma =2$, $\protect\kappa =0.1$, moving at velocity $\left(
v_{y}=0.1,v_{x}=0\right) $. (c,d) The evolution of the same soliton,
simulated in the framework of Eqs. (\protect\ref{phipsi}) (written in the
quiescent coordinates, rather than moving ones), corroborates its stability.
}
\label{MM_moving}
\end{figure}

Following Ref. \cite{Ben-Li}, localized states which move steadily at
velocity $\mathbf{v}=(v_{x},v_{y})$ can be looked for in the form of $\phi
_{\pm }=\phi _{\pm }(x-v_{x}t,y-v_{y}t,t)$ and $\psi _{\pm }=\psi _{\pm
}(x-v_{x}t,y-v_{y}t,t)$. The substitution of this in Eq. (\ref{phipsi})
leads to the equations written in the moving reference frame,

\begin{eqnarray}
i\frac{\partial \phi _{+}}{\partial t}-i\left( \mathbf{v}\cdot \nabla
\right) \phi _{+} &=&-\frac{1}{2}\nabla ^{2}\phi _{+}-\left( |\phi
_{+}|^{2}+\gamma |\phi _{-}|^{2}\right) \phi _{+}+\lambda \left( \frac{%
\partial \phi _{-}}{\partial x}-i\frac{\partial \phi _{-}}{\partial y}%
\right) -\kappa \psi _{+},  \notag \\
i\frac{\partial \phi _{-}}{\partial t}-i\left( \mathbf{v}\cdot \nabla
\right) \phi _{-} &=&-\frac{1}{2}\nabla ^{2}\phi _{-}-\left( |\phi
_{-}|^{2}+\gamma |\phi _{+}|^{2}\right) \phi _{-}-\lambda \left( \frac{%
\partial \phi _{+}}{\partial x}+i\frac{\partial \phi _{+}}{\partial y}%
\right) -\kappa \psi _{-},  \notag \\
&&  \label{phipsi2} \\
i\frac{\partial \psi _{+}}{\partial t}-i\left( \mathbf{v}\cdot \nabla
\right) \psi _{+} &=&-\frac{1}{2}\nabla ^{2}\psi _{+}-\left( |\psi
_{+}|^{2}+\gamma |\psi _{-}|^{2}\right) \psi _{+}+\lambda \left( \frac{%
\partial \psi _{-}}{\partial x}-i\frac{\partial \psi _{-}}{\partial y}%
\right) -\kappa \phi _{+},  \notag \\
i\frac{\partial \psi _{-}}{\partial t}-i\left( \mathbf{v}\cdot \nabla
\right) \psi _{-} &=&-\frac{1}{2}\nabla ^{2}\psi _{-}-\left( |\psi
_{-}|^{2}+\gamma |\psi _{+}|^{2}\right) \psi _{-}-\lambda \left( \frac{%
\partial \psi _{+}}{\partial x}+i\frac{\partial \psi _{+}}{\partial y}%
\right) -\kappa \phi _{-},  \notag
\end{eqnarray}%
where $x$ and $y$ actually stand for $x-v_{x}t$ and $y-v_{y}t$. Stationary
solutions to these equations have been obtained by means of the same AITEM
technique which was also used, as mentioned above, for producing quiescent
solitons. As well as in Ref. \cite{Ben-Li}, moving MMs are robust states,
surviving up to relatively high speeds, before they undergo delocalization
at a critical value of the speed. In particular, setting, as above, $\gamma
=2$ in Eqs. (\ref{phipsi2}), a typical example of a stable moving symmetric
MM soliton is displayed in Fig. \ref{MM_moving}, for
\begin{equation}
(N,\kappa )=(3,0.1)  \label{Nkappa}
\end{equation}%
and $\left( v_{x},v_{y}\right) =\left( 0,0.1\right) $. Panels (c,d) of the
figure corroborate the stability of the moving soliton, whose stationary
shape is displayed in panels (a,b).

Note that the quiescent MM soliton with the same values of the parameters as
given by Eq. (\ref{Nkappa}), is unstable against the SSB, as seen in Fig. %
\ref{MM_osci}, unlike the stable moving soliton in Fig. \ref{MM_moving}.
This means that motion helps to stabilize the symmetric MMs against the
symmetry breaking. Effects of the motion are summarized in Fig. \ref%
{A_velocity1}(a), which shows the amplitude of the moving symmetric MM,
defined as
\begin{equation}
A=\sqrt{|\phi _{+}(0,0)|^{2}+|\phi _{-}(0,0)|^{2}}\equiv \sqrt{|\psi
_{+}(0,0)|^{2}+|\psi _{-}(0,0)|^{2}}  \label{A}
\end{equation}%
(in the moving coordinates), vs. $v_{y}$. First, we note that, as said
above, the symmetric MM soliton is unstable at $v_{y}=0$, and the increase
of the velocity leads to the stabilization at $v_{y}\approx 0.03$. Further,
amplitude (\ref{A}) \emph{monotonically decreases} with the growth of the
velocity, and the soliton vanishes, through delocalization, at a critical
speed,
\begin{equation}
\left( v_{y}\right) _{\max }^{\mathrm{(MM)}}\approx 0.18.  \label{vymax}
\end{equation}

The conclusion concerning the robustness of the moving MM solitons is in
qualitative agreement with the finding displayed above in Fig. \ref{MM_osci}%
, which demonstrates that the SSB transforms unstable MMs into stably
drifting breathers.

\begin{figure}[tbp]
\centering{\label{fig15a} \includegraphics[scale=0.2]{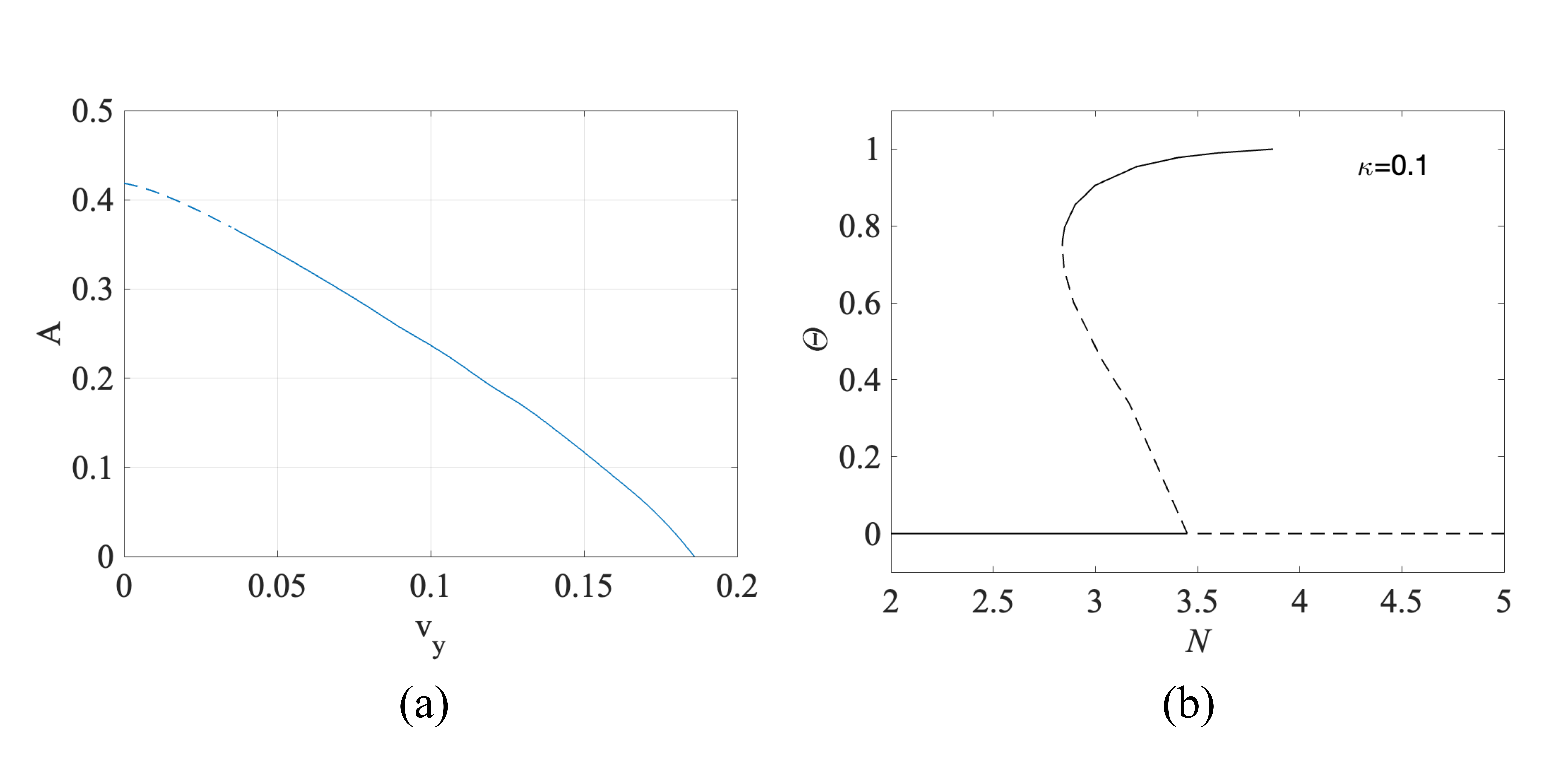}}
\caption{(Color online) (a) The amplitude of the moving symmetric MM soliton
as a function of its velocity, $v_{y}$, for the fixed total norm and
inter-layer coupling constant, $N=3$ and $\protect\kappa =0.1$. (b) The
symmetry-breaking bifurcation of the moving 2D two-layer MMs, for the same $%
\protect\kappa =0.1$ and fixed velocity $v_{y}=0.1$ . In both panels (a) and
(b), solid and dashed curves designate stable and unstable solutions,
respectively. These plots are produced for $\protect\gamma =2$ in Eqs. (%
\protect\ref{phipsi2}).}
\label{A_velocity1}
\end{figure}

\begin{figure}[tbp]
\centering{\label{fig16a} \includegraphics[scale=0.1]{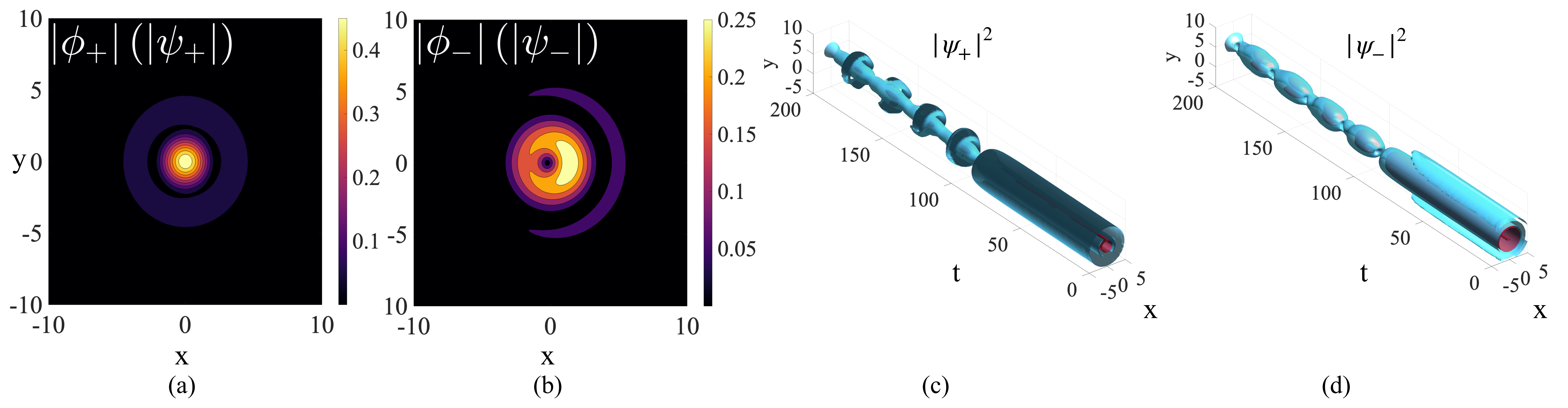}}
\caption{(Color online) Contour plots of $|\protect\psi _{+}(x,y)|$ (or $|%
\protect\phi _{+}(x,y)|$) (a) and $|\protect\psi _{-}(x,y)|$ (or $|\protect%
\phi _{-}(x,y)|$) (b) of a stable symmetric SV soliton with $N=5$ and $%
\protect\kappa =0.1$, moving at velocity $v_{y}=0.005$ (while $v_{x}=0$).
(c,d) The evolution of an unstable symmetric SV soliton with $(N,\protect%
\kappa ,v_{y})=(5.6,0.1,0.005)$, produced by simulations of Eq. (\protect\ref%
{phipsi}). It shows the onset of the spontaneous symmetry breaking with
concomitant oscillations. The unstable soliton keeps traveling in the $y$
direction at the initial velocity, $v_{y}=0.005$.}
\label{SV_moving}
\end{figure}

The SSB of the moving MM solitons, with fixed velocity $v_{y}=0.1$, is
summarized in Fig. \ref{A_velocity1}(b) by means of the respective
symmetry-breaking diagram, which shows asymmetry parameter (\ref{ratio}) vs.
the total norm, $N$, defined as above [according to Eq. (\ref{Norm})]. In
this case, the SSB point is $N_{\mathrm{cr}}\approx 3.5$, which is
essentially larger than its counterpart, $N_{\mathrm{cr}}\approx 2.8$, found
above for the quiescent MM solitons, cf. Fig. \ref{MM_theta_kappa}(b). This
finding corroborates the above conclusion, that the motion helps to
stabilize the MM solitons against the SSB. It is worthy to note that the SSB
bifurcation for the moving solitons always remains subcritical.

Although Eqs. (\ref{phipsi2}) seem anisotropic in the $\left( x,y\right) $
plane, the numerical analysis of the MM solitons moving in the $x$ direction
produces the same results, as concerns the dependences between the amplitude
and velocity, and the SSB diagram, as shown in Fig. \ref{A_velocity1} for
the solitons moving along the $y$ axis. Thus, the system is actually
isotropic for moving solitons (this point was not considered in Ref. \cite%
{Ben-Li}).

Unlike the MMs, moving SV solitons are fragile, suffering delocalization at
small values of the velocity (for the single-layer system, this conclusion
was made in Ref. \cite{Ben-Li}). For example, at parameters $\gamma =0$ and $%
\kappa =0.1$, the moving SVs with norms $N=5$ and $7.4$ exist up to the
critical speeds $(v_{y})_{\max }^{\mathrm{(SV)}}(N=5)\approx 0.01$ and $%
(v_{y})_{\max }^{\mathrm{(SV)}}(N=7.4)\approx 0.03$, respectively, cf. the
much larger critical speed (\ref{vymax}) for the moving MMs. At $\left\vert
v_{y}\right\vert >(v_{y})_{\max }^{\mathrm{(SV)}}$, numerical solution to
Eqs. (\ref{phipsi2}), produced by AITEM, does not demonstrate
delocalization. Instead, the moving SVs convert into MM states, instead of
SVs. This conclusion also agrees with the findings for the single-layer
system, reported in Ref. \cite{Ben-Li}.

A typical example of a stably moving symmetric SV is presented in Figs. \ref%
{SV_moving}(a,b) for $(N,\kappa ,\gamma )=(5,0.1,0)$ and $v_{y}=0.005$.
Panels (c) and (d) of the figure display the evolution of an initially
symmetric SV with $(N,\kappa ,\gamma )=(5.6,0.1,0)$ and the same speed, $%
v_{y}=0.005$, which is unstable against SSB, due to the slightly larger
value of the total norm. It is seen that the dynamical symmetry breaking
sets in along with concomitant oscillations. Generally, this is similar to
the SSB-induced instability in the quiescent SVs, cf. Fig. \ref{SV_DR1}.

It is relevant to note that the shape of the SV soliton moving along the $y$
axis, which is displayed in Fig. \ref{SV_moving}, keeps spatial symmetry
with respect to the $x$ axis, and features conspicuous asymmetry with
respect to $y$ (the same asymmetry, but in a weaker form, is observed for
moving SV solitons in Fig. \ref{SV_moving}, as well as in Ref. \cite{Ben-Li}%
). The distortion of the axial symmetry of the quiescent SV solitons (see
Fig. \ref{2DASVs}) under the action of motion is a manifestation of the
Magnus effect for vortex solitons, cf. Rev. \cite{Magnus}. 

\section{Conclusion}

The objective of this work is to extend the analysis of 2D solitons
supported by the interplay of the cubic self- and cross-attraction and
linear SOC (spin-orbit coupling) in the model of the binary BEC. A
surprising result, first reported in Ref. \cite{Ben-Li}, was that this
system gives rise to absolutely stable 2D solitons of the SV (semi-vortex),
MM (mixed-mode), and CS (composite-soliton) types, the latter one only
existing in the case of the Manakov's nonlinearity. These solitons furnish
the system's ground state, which does not exist in the absence of the SOC
terms (in that case, the collapsing regime formally plays the role of the
ground state). In the present work, we aimed to consider the double-layer
setting, in which two identical nonlinear systems stabilized by SOC are
linearly coupled by tunneling of atoms. Stationary solutions of the
two-layer model were produced by means of AITEM (accelerated imaginary-time
evolution method), and their stability was tested by means of direct
real-time simulations.

It has been concluded that the 2D solitons of the SV, MM, and CS types,
which are initially symmetric with respect to the tunnel-coupled layers, are
subject to the SSB (spontaneous-symmetry-breaking) bifurcation, which is
always of the subcritical type. Branches of stable asymmetric 2D solitons,
produced by the SSB bifurcation, exist up to the value of the total norm
equal to that of the TSs (Townes solitons), as the collapse takes place at
norms exceeding the TS value. This picture is valid up to a certain largest
value of the inter-layer coupling constant; above that value, the
bifurcation does not take place, because the collapse commences prior to it.
The evolution of initially symmetric 2D solitons of the SV and MM types,
which are destabilized by the SSB, features dynamical symmetry breaking,
along with the establishment of intrinsic oscillations of the soliton, or
quick development of the collapse, if the total norm is large enough to
drive the collapse. In the case of MMs, the breather produced by the
symmetry-breaking instability features spontaneous drift, in addition to the
intrinsic oscillations.

Moving 2D solitons were investigated in the traveling reference frame. This
is a relevant problem because the SOC breaks the system's Galilean
invariance. It has been found that moving MMs are robust 2D solitons, which
exist up to the critical value of the velocity, above which they suffer
delocalization. Traveling SVs are fragile states, for which the critical
speed is very small. Above it, the solitons do not decay, but spontaneously
rearrange themselves into moving MM state. This phenomenon was also observed
in a single-layer model in Ref. \cite{Ben-Li}.

To extend the present work, it may be relevant to study interactions between
the stable 2D solitons. This problem should be especially interesting for
interactions between stable asymmetric solitons. Additionally, it is also
interesting to consider initially asymmetric pair of layers with unequal SOC
strengths $\lambda $ in them. Consideration of the nonlinear interaction
between the layers, in addition to the linear coupling, may be relevant too.

\section*{Acknowledgments}

We appreciate valuable discussions with H. Sakaguchi. This work was
supported, in part, by the Israel Science Foundation through grant No.
1695/22, and in part, by Key Research Projects of General Colleges in
Guangdong Province through grant No. 2019KZDXM001. Z.C. acknowledges a
fellowship provided by the Helen Diller quantum center at the Technion
(Haifa, Israel).

\end{document}